\documentclass[aps,pra,twocolumn,groupedaddress,showpacs]{revtex4}

\usepackage{graphicx}
\begin{document}


\title{Coherent control with shaped femtosecond laser pulses applied to ultracold molecules}


\author{Wenzel Salzmann}
\author{Ulrich Poschinger}
\author{Roland Wester}
\author{Matthias Weidem\"uller}
\email[]{Corresponding_author:
m.weidemueller@physik.uni-freiburg.de}

\affiliation{Physikalisches Institut, Universit\"at Freiburg, Hermann Herder
Str. 3, D-79104 Freiburg i. Br.}

\author{Andrea Merli},
\author{Stefan M. Weber}
\author{Franziska Sauer}
\author{Mateusz Plewicki}
\author{Fabian Weise}
\author{Aldo Mirabal Esparza}
\author{Ludger W\"oste}
\author{Albrecht Lindinger}
\email[]{Corresponding_author: lindin@physik.fu-berlin.de}
\affiliation{Institut f\"ur Experimentalphysik, Freie
Universit\"at Berlin, Arnimallee 14, D-14195 Berlin, Germany}

\begin{abstract}
We report on coherent control of excitation processes of translationally
ultracold rubidium dimers in a magneto-optical trap by using shaped
femtosecond laser pulses. Evolution strategies are applied in a feedback loop
in order to optimize the photoexcitation of the Rb$_2$ molecules, which
subsequently undergo ionization or fragmentation. A superior performance of
the resulting pulses compared to unshaped pulses of the same pulse energy is
obtained by distributing the energy among specific spectral components. The
demonstration of coherent control to ultracold ensembles opens a path to
actively influence fundamental photo-induced processes in molecular quantum
gases.
\end{abstract}

\pacs{32.80.Qk, 33.80.-b, 82.53.-k}

\maketitle

\section{Introduction}

The interest to study interactions in ultracold molecular
gases~\cite{doyle2004} is propelling various research programs to produce and
trap dense ultracold molecular ensembles
\cite{fioretti1998,weinstein1998:nat,bethlem1999:prl,jochim2003:sci,greiner2003:nat,zwierlein2003:prl}.
The goals and perspectives in this field range from quantum computation
\cite{demille2002} and quantum scattering \cite{chin2005} to ultrahigh
precision spectroscopy \cite{hinds2002} and coherent ultracold chemistry
\cite{heinzen2000}.
The manipulation of molecular wavepacket dynamics with shaped femtosecond
pulses may open exciting possibilities to this field. In recent years, the
control of molecular processes by shaped fs-laser pulses through application
of evolution strategies in a feedback loop has attained considerable
success. Since it was proposed \cite{judson1992}, coherent control with shaped
femtosecond laser pulses has been applied to a large variety of experiments,
i.e. selective fragmentation \cite{assion1998}, population transfer
\cite{bareen1997}, ionization \cite{vajda2001}, and high harmonic generation
\cite{bartels2000}.  By using an iterative process, the method enables the
discovery of an optimal laser pulse shape which drives the system towards a
desired target state.
It even became possible to decipher the molecular dynamics underlying the
laser-molecule interaction by analyzing the optimal pulse shapes
\cite{vajda2001,daniel2003:sci}.

Theoretical work on photoassociation with pulsed lasers has started several
years ago \cite{machholm1994} and first experiments with thermal mercury atoms
in a gas cell \cite{marvet1995} and with sodium atoms in a magneto-optical
trap \cite{fatemi2001} have been performed. More recently, optimal control
calculations were done proposing efficient ultracold molecule formation
\cite{vala2000} and vibrational cooling \cite{koch2004} by appropriately
shaped laser pulses. However, up to now there are no experimental results on
the application of feedback-controlled optimization to photoassociation or
manipulation of ultracold molecules.
In this article we present first experimental steps towards this goal by
investigating the interaction of femtosecond laser pulses with ultracold
ground state molecules in view of a pulsed photoassociation experiment. The
low temperature of the atomic and molecular sample, while not directly
influencing the interaction with the femtosecond pulses, leads to molecules in
very high lying vibrational levels and allows us to study molecules under the
same conditions of typical photoassociation experiments. In this work we
demonstrate in particular the enhancement of the electronic excitation and
subsequent fragmentation of ultracold molecules from a magneto-optical trap
using shaped femtosecond laser pulses, which were adaptively optimized by an
evolutionary algorithm.

\section{Experimental setup}

\begin{figure}[!tbp]
\begin{center}
\includegraphics[width=\columnwidth]{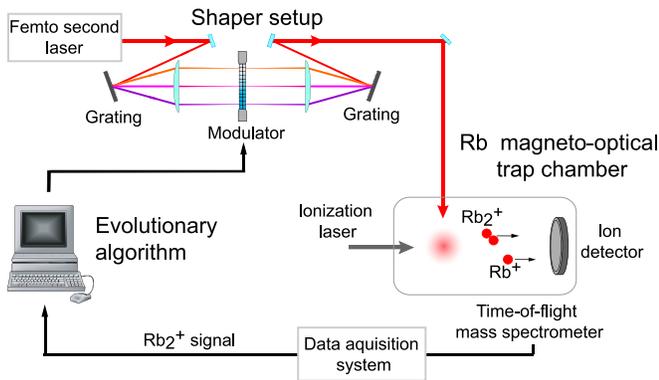}
\caption{Experimental setup for iterative closed-loop maximization of
ultracold Rb$_2$ excitation from the ground electronic singlet or triplet
states by shaped femtosecond laser pulses. The ultracold molecules are formed
in a magneto-optically trapped gas of rubidium atoms. The laser pulse shapes
represent individuals of an evolutionary algorithm. Their fitness is evaluated
by the reduction of the Rb$_2^+$ signal resulting from resonantly enhanced
photoionization of the Rb$_2$ molecules.} \label{setup}
\end{center}
\end{figure}

In our experiment, schematically depicted in Fig. \ref{setup}, about $10^7$
$^{85}$Rb atoms are captured in a magneto-optical trap (MOT) at a density of
$10^{10}$\,atoms/cm$^3$ and a temperature of $100\,\mu$ K.
Diatomic rubidium molecules continuously form in the magneto-optical trap due
to either three body collisions or light assisted two-body collisions of
trapped Rb atoms. They populate predominantly the highest vibrational states
below the dissociation limit in the triplet electronic ground state
\cite{gabbanini2000,helm2004}. These Rb$_2$ molecules, which are not trapped
by the trap magnetic field, are detected in this experiment via resonant two
photon ionization and time-of-flight mass analysis. The ionization laser is a
15\,Hz Nd:YAG pumped dye laser at wavelengths between 600 and 610\,nm, a
spectral width of 0.5\,cm$^{-1}$ and a pulse energy of 20\,mJ. In the steady
state of molecule formation and loss a maximum count rate of 0.5 Rb$_2^+$
molecular ions per laser pulse is observed, measured at a wavelength of
602.5\,nm.

The femtosecond (fs) laser pulses are generated in a Ti:Sapphire oscillator
(Tsunami; Spectra Physics) that provides pulses of 120\,fs duration (FWHM) at
a rate of 80\,MHz, a spectral width of $\Delta \lambda = 10$\,nm and an energy
of up to 18\,nJ per pulse.  Low pulse energies ensure that the laser-molecule
interaction is well described in a perturbative picture and no high intensity
effects have to be considered. To modify the spectral components of the pulses
we use a pulse shaper that allows independent phase and amplitude modulation
\cite{Wefers95}. It consists of a liquid crystal modulator (CRI; SLM-256)
\cite{Weiner92} with 2x128 pixels, placed in the Fourier plane of a double
grating zero dispersion compressor. A lens focuses the beam to a spot of
150\,$\mu$m diameter at the center of the trap, illuminating about 10\% of the
cloud volume. For the experiments the fs-laser is tuned in the range between
780 and 820\,nm.


\section{Laser-induced loss}

Transform-limited fs-pulses with wavelengths near the Rb resonance
lines and focused into the MOT are found to strongly interact
with the trapped atoms, as observed through a significant decrease
of MOT fluorescence. At small pulse energies, the trap loading
rate outside the femtosecond beam can partly compensate for the
losses in the focus, resulting in an area of reduced fluorescence
predominantly where the beam is passing, whereas for high energies
the trap is completely depleted. We attribute this to photon
scattering from the femtosecond pulses, which causes an outward
directed light force on the atoms, leading to trap loss. The
effect is significantly stronger when the D2 atomic resonance at
780\,nm was part of the pulse spectrum than for the D1 resonance
at 795\,nm, reflecting the higher transition dipole moment of the
D2 line. A narrow slit, representing a bandpass of 0.3 nm width,
is scanned through the Fourier plane of the pulse shaper, which
showed that only components resonant with the D1 and D2 atomic
transitions are responsible for this effect. Singly-charged
rubidium ions have been detected when the central pulse frequency
was resonant with one of the atomic transitions, indicating that
resonant three-photon ionization contributes to the loss of atoms
from the MOT. To study the laser pulse interaction with rubidium
molecules, the atomic resonance components were removed from the
pulse spectrum by a notch filter, realized by a physical block in
the shaper's Fourier plane. In this way atomic losses from the MOT
could be reduced below the detection threshold.

\begin{figure}[!tbp]
\begin{center}
\includegraphics[width=\columnwidth]{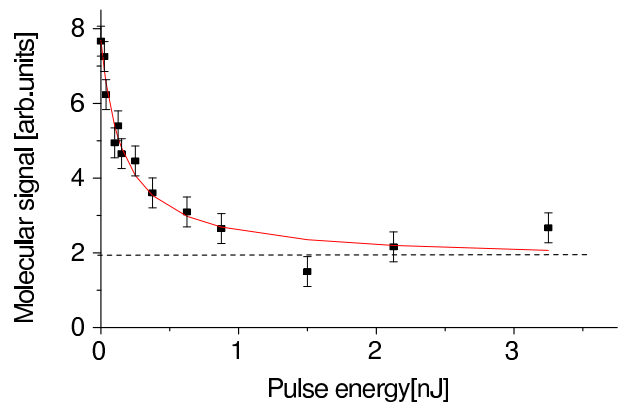}
\caption{Reduction of the Rb$_2^+$ molecular ion signal as a function of
energy of transform limited femtosecond pulses. The pulses have a central
wavelength of 800\,nm and 10\,nm FWHM. The D1 atomic resonance at 795\,nm is
filtered out of the pulse. The ionization laser for REMPI is set to
602.5\,nm. The solid curve represents a fit to a single photon excitation
model as described in the text.}
\label{moldiss602}
\end{center}
\end{figure}

The rubidium dimers interact with the fs-laser pulses over the entire
accessible range of central wavelengths from 780\,nm to 820\,nm.
As shown in Fig. \ref{moldiss602}, the molecular signal decreases rapidly at
small pulse energies and levels off to 25\% at a pulse energy of 0.6\,nJ. As
only molecules in the electronic ground state are detected, the signal
reduction can be attributed to excitation by the fs-pulses. The process can be
modeled by a simple rate equation for the number of detected ground state
molecules:
\begin{equation}
\frac{dN_{\rm Mol}}{dt} = R_{\rm form}- (\Gamma_{\rm fs} + \Gamma_{\rm
loss}) N_{\rm Mol}
\end{equation}
where $R_{\rm form}$ is the production rate of molecules from trapped atoms,
$\Gamma_{\rm fs}$ is the rate of molecular loss induced by interaction
with the femtosecond laser, and $\Gamma_{\rm loss}$ the molecular loss rate
induced by other processes such as rest gas collisions or gravitation
accelerating the molecules out of the detection volume. In the steady state
$dN_{\rm Mol}/dt = 0$ the number of molecules is given by
\begin{equation}
N_{\rm Mol} = \frac{R_{\rm form}}{\Gamma_{\rm loss}+ \Gamma_{\rm fs}}
\end{equation}
The curve in fig.\,\ref{moldiss602} represents a fit to the data assuming a
linear dependence of $\Gamma_{\rm fs}$ on the pulse energy of the laser,
i.e. $\Gamma_{\rm fs} \propto P_{\rm fs}$. The assumption of a quadratic or
higher order dependence of the excitation rate on pulse energy does not fit
the data. This indicates that, in the regime of pulse energies employed in the
experiment, the interaction with the molecules has the character of an
effective one-photon excitation \cite{ban2005}.

\begin{figure}[!tbp]
\begin{center}
\includegraphics[width=\columnwidth]{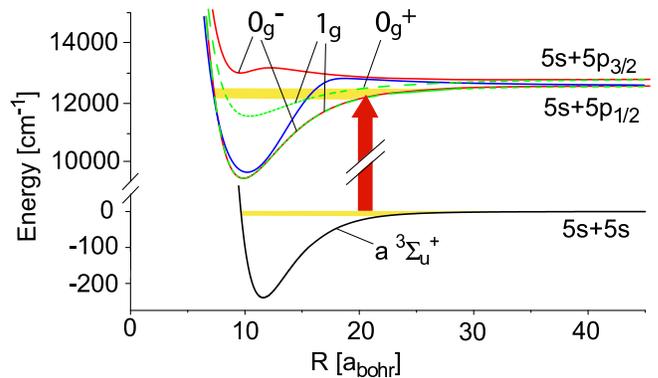}
\caption{Potential curves of the rubidium dimer including spin-orbit
interaction. Initially, the molecules are expected to populate the highest
levels in the a$^3\Sigma_u^+$ \cite{gabbanini2000,helm2004}.  The molecular
excitation by the femtosecond pulses is indicated by the arrow. The shaded
areas show the initial and final distribution of molecular vibrational
states.} \label{potentials}
\end{center}
\end{figure}

According to \cite{gabbanini2000}, the molecules in the MOT initially populate
the highest levels in the a$^3\Sigma_u^+$ state. Due to selection rules and
Franck-Condon factors, they are preferably excited to the 0$_g^-$ and 1$_g$
5s5p$_{1/2}$ states (see Fig.\ \ref{potentials}).
Emission back into the electronic ground state could only lead to signal
reduction if the vibrational level population moves out of the excitation
window of the ionization laser. However, scans of the detection laser with and
without fs-beam show reduced but qualitatively similar spectra which should
not be the case for a vibrational redistribution. Instead
it can be expected that excited molecules absorb further photons, so the whole
process of molecular loss can be regarded as a resonance enhanced multi-photon
excitation, followed by dissociation, predissociation or ionization. This
happens either within one pulse, or, as the laser repetition rate is
comparable to the lifetime of the first excited state, it occurs in the
subsequent pulse. At high energies all molecules in the laser focus are
excited or dissociated and the residual signal in Fig.\ \ref{moldiss602} is
due to molecules which did not interact with the femtosecond laser. This shows
that most of the molecules are produced within a small volume inside the MOT
which is consistent with the picture that they form at the MOT center where
the atom number density is at its maximum \cite{townsend1995}. At high
energies all molecules in the laser focus are excited or dissociated and the
residual signal in Fig.\ \ref{moldiss602} is due to molecules which did not
interact with the femtosecond laser, indicated by the dashed line. This shows
that most of the molecules are produced within a small volume inside the MOT
which is consistent with the picture that they form at the MOT center where
the atom number density is at its maximum \cite{townsend1995}.

\section{Application of coherent control}

In order to demonstrate the practical applicability of coherent control
concepts to ultracold molecules, the Rb$_2^+$ signal acts as an input for a
self-learning optimization algorithm which autonomously programs the pulse
shaper in a closed loop experiment. The algorithm is based on evolution
strategies and is described in detail in \cite{bartelt2001}. Because of the
small molecular ion count rate and hence the low signal-to-noise ratio the
signal is averaged over 128 dye laser pulses for each individual of the
algorithm. To reduce the search space for the
learning algorithm we chose a mixed scheme of parametric amplitude and free
phase optimization where the algorithm tries to find the optimal pulse shape
under the restriction that only a few sharp spectral peaks contribute to the
pulse shape. During an optimization the parameters of these peaks, their
spectral positions and amplitudes, are altered to find the most efficient
excitation pulse. Moreover, the phase was optimized freely in order to allow a
temporal modulation of the pulse. The evolutionary algorithm administrates 31
individuals, each representing the pulse shaper parameters to produce pulses
consisting of up to eleven Gaussian peaks of 7 cm$^{-1}$ FWHM.

\begin{figure}[!tbp]
\begin{center}
\includegraphics[width=\columnwidth]{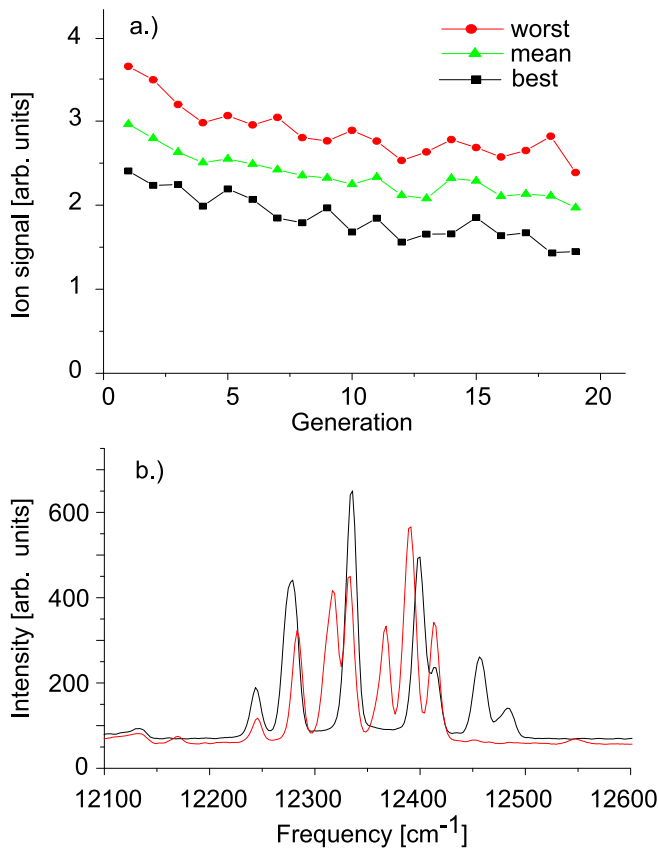}
\caption{(a) Molecular ion signal resulting from the best, the
worst and the mean individual of the population for each
generation during a closed loop experiment. (b) Femtosecond laser
pulse spectrum of the final best individuals of two successive
optimization runs under equal conditions with similar final
optimization result.} \label{Verlauf}
\end{center}
\end{figure}

The adaptive algorithm was applied to manipulate the excitation pulses with
the aim to minimize the molecular signal from the MOT.  For each iteration the
ion signals corresponding to the best and worst individuals are protocoled
together with the mean fitness of the whole generation. As depicted in Fig.\
\ref{Verlauf}(a), all three signals decrease during the particulate
optimization to about 70\% of the initial value after 20 iterations. The
spectra of the final best individuals of two successive runs shown in Fig.
\ref{Verlauf}(b) display several peaks which coincide in some but not all
spectral positions. The frequency span of the fs-pulse supports our assignment
of excitation to the 0$_g^-$ and 1$_g$ 5s5p$_{1/2}$ states (see Fig.\
\ref{potentials}). By comparing the excitation yield of the best individuals
with transform-limited pulses of the same energy it is observed that the
optimized pulse excites the molecules on average 25\% more efficiently, which
demonstrates the feasibility and potential of adaptive control.


We attribute the observed excitation enhancement to an increased spectral
intensity at particular molecular resonances found by the evolutionary
algorithm. Starting from a narrow band in the a$^3\Sigma_u^+$ state
\cite{gabbanini2000,helm2004}, molecules are excited into bound states below
the D1 resonance. By shifting the peak positions, the algorithm finds
transition frequencies from this band to certain vibrational states, thereby
sharing the pulse energy more efficiently than a broad Gaussian pulse. The
algorithm therefore has a large number of possible solutions to choose from
and so the final pulse shapes after an optimization are not identical. In the
spectral region between 12000 and 12500 cm$^{-1}$, the vibrational level
separation is about 10 cm$^{-1}$ in the 0$_g^-$ and 1$_g$ 5s5p$_{1/2}$ states,
respectively. The high density of states also explains the limited potential
of the optimization because the optimization factor depends on the chosen
peakwidth which is limited by the shaper resolution. The Franck-Condon factors
may also be relevant for the excitation process since they differ for
different vibronic transitions and favor particular frequencies which are
enhanced in the experimentally acquired spectra. Yet, as the initial ground
state population distribution in the vibrational states is not known
accurately, no quantitative treatment or assignment can be made.

\section{Conclusions} 

In this article we demonstrate the application of iterative adaptive control
techniques to the manipulation of ultracold molecules. The minimization of
molecular signal from a rubidium magneto-optical trap results in pulse shapes
that significantly excite more Rb$_2$ molecules than transform-limited pulses
of the same energy. The resulting pulse spectra are not unambiguous because of
the large number of possible optimal solutions within the experimental
accuracy. For future applications of shaped fs-pulses in photoassociation
experiments, the dissociative effect of the pulses has to be suppressed. On
the one hand, photoassociated cold molecules may be dissociated by subsequent
femtosecond pulses due to the high repetition rate of the femtosecond laser
system. Therefore, photoassociated molecules have to be detected directly
after their formation by increasing the repetition rate of the detection
system, ideally matching the rate of the photoassociating beam. On the other
hand, the excitation of ground state molecules may be coherently suppressed
while enhancing the molecule formation through actively shaping the
pulses. For this purpose, we currently theoretically investigate realistic
scenarios for femtosecond photoassociation of cold molecules by numerically
simulating the dynamical behavior of an atom pair interacting with an
arbitrarily shaped electromagnetic field. It is also envisioned to use
specifically shaped femtosecond laser pulses for the cooling of vibrational
molecular excitations by pump-dump processes via an intermediate excited
state. This approach could be applied to produce quantum-degenerate molecular
gases in arbitrary vibrational states or superpositions of states.

\begin{acknowledgments}

We thank C. Koch, R. Kosloff, B. Sch{\"a}fer-Bung and V.
Bona\v{c}i{\'c}-Kouteck{\'y} for theoretical support and many stimulating
discussions. This work was supported by the Deutsche Forschungsgemeinschaft in
the frame of the Sonderforschungsbereich 450 and the Schwerpunktprogramm
1116. F. Sauer acknowledges the Studienstiftung des deutschen Volkes, A. Merli
thanks the Cusanuswerk.

\end{acknowledgments}

\end{document}